\documentclass[fleqn,usenatbib]{mnras}

\usepackage{newtxtext,newtxmath}

\usepackage[T1]{fontenc}

\DeclareRobustCommand{\VAN}[3]{#2}
\let\VANthebibliography\thebibliography
\def\thebibliography{\DeclareRobustCommand{\VAN}[3]{##3}\VANthebibliography}

\usepackage{graphicx}	
\usepackage{newtxtext,newtxmath}
\usepackage{booktabs}
\usepackage{tabularx}
\usepackage{makecell}
\usepackage{multicol}
\usepackage{changepage}
\usepackage{color, soul}

\title[Deep Learning for photo-z]{Investigating Deep Learning Methods for Obtaining Photometric Redshift Estimations from Images}

\author[B. Henghes et al.]{
Ben Henghes,$^{1}$\thanks{E-mail: ben.henghes.13@ucl.ac.uk}
Connor Pettitt,$^{2}$
Jeyan Thiyagalingam,$^{2}$\thanks{E-mail: t.jeyan@stfc.ac.uk}
Tony Hey,$^{2}$
and Ofer Lahav,$^{1}$
\\
$^{1}$ Department of Physics \& Astronomy, University College London, Gower Street, London, WC1E 6BT, UK \\
$^{2}$Scientific Computing Department, Rutherford Appleton Laboratory, Science and Technology Facilities Council (STFC), Harwell Campus, Didcot, OX11 0QX, UK 
}

\date{Accepted XXX. Received YYY; in original form ZZZ}

\pubyear{2021}

\begin{document}
\label{firstpage}
\pagerange{\pageref{firstpage}--\pageref{lastpage}}
\maketitle

\begin{abstract}
Knowing the redshift of galaxies is one of the first requirements of many cosmological experiments, and as it’s impossible to perform spectroscopy for every galaxy being observed, photometric redshift (photo-$z$) estimations are still of particular interest. Here, we investigate different deep learning methods for obtaining photo-$z$ estimates directly from images, comparing these with `traditional' machine learning algorithms which make use of magnitudes retrieved through photometry. As well as testing a convolutional neural network (CNN) and inception-module CNN, we introduce a novel mixed-input model which allows for both images and magnitude data to be used in the same model as a way of further improving the estimated redshifts. We also perform benchmarking as a way of demonstrating the performance and scalability of the different algorithms. The data used in the study comes entirely from the Sloan Digital Sky Survey (SDSS) from which 1 million galaxies were used, each having 5-filter (ugriz) images with complete photometry and a spectroscopic redshift which was taken as the ground truth. The mixed-input inception CNN achieved a mean squared error ($MSE$) $=0.009$, which was a significant improvement ($30\%$) over the traditional Random Forest (RF), and the model performed even better at lower redshifts achieving a $MSE=0.0007$ (a $50\%$ improvement over the RF) in the range of $z < 0.3$. This method could be hugely beneficial to upcoming surveys such as the Vera C. Rubin Observatory's Legacy Survey of Space and Time (LSST) which will require vast numbers of photo-$z$ estimates produced as quickly and accurately as possible. 
\end{abstract}




\section{Introduction}

In the past decade the number of galaxies observed by large sky surveys has been rapidly increasing with hundreds of millions of galaxies being imaged \citep{alam2015eleventh, drlica2018dark, kids_dr}. This ever growing number is set to rise even faster with upcoming surveys such as the Vera C. Rubin Observatory’s Legacy Survey of Space and Time (LSST) \citep{tyson2003lsst} and the Roman Space Telescope (formerly WFIRST - the Wide-Field Infrared Survey Telescope) \citep{spergel2015wide} which will observe many billions of objects. For most cosmological studies in which galaxies are used the redshift is a key property that is required; however, despite spectroscopic surveys such as the Dark Energy Spectroscopic Instrument (DESI) \citep{flaugher2014dark, desi_2}, only a very small fraction of galaxies have an associated spectroscopic redshift. Instead, photometric redshift (photo-$z$) estimations are necessary. 

These photo-$z$ estimates can be obtained using two different methods (or a combination of both): template fitting, or machine learning (ML). The template fitting method uses templates of the spectral energy distribution (SED), and by fitting the observed SED of the galaxy to the template, its redshift can be inferred \citep{bolzonella2000photometric}. ML methods instead use a large training set of galaxies with labelled, `true' values for the redshift (the spectroscopic redshift) and learn a mapping from the features of the galaxy data to their redshifts \citep{collister2004annz}. For traditional ML techniques, such as Random Forests (RF) or k-Nearest Neighbours (kNN), these features are taken from the photometry, giving magnitudes in different filters which can be combined into colours. However, for deep learning methods the image itself can be used as the input, with the pixel values being akin to features \citep{hoyle2016measuring}.

There are benefits to each method with template fitting being an inherently physical model, and by making use of SED templates which are full spectra, they can be shifted to any redshift and allow for redshift estimates to be obtained in ranges without large spectroscopic datasets \citep{benitez2000bayesian}. ML methods on the other hand require a representative training sample in the same redshift distribution as the targets and as a result are only valid in that range. Despite this limitation, ML has been increasingly implemented as a faster method which is able to produce very accurate photo-$z$ estimates where there is a sufficiently large training set \citep{abdalla2011comparison}. 

Recently great strides have been made in the field of deep learning, aided by improving computer architectures and faster graphics processing units (GPUs), far more difficult tasks can now be performed using much larger datasets \citep{kirk2007nvidia}. In industry many companies have been taking advantage of these methods for tasks which range from creating outfits for fast-fashion \citep{bettaney2019fashion}, to self driving vehicles \citep{bojarski2016end}. In astronomy surveys such as the Dark Energy Survey (DES) \citep{DES_more}, the Kilo-Degree Survey (KiDS) \citep{de2013kilo}, Euclid \citep{amendola2018cosmology}, Hyper Suprime-Cam (HSC) \citep{aihara2018hyper}, LSST \citep{Ivezi__2019}, the Roman Space Telescope \citep{spergel2015wide}, and the Square Kilometer Array (SKA) \citep{dewdney2009square}, will be producing petabytes of data and machine learning provides a viable solution to the otherwise unimaginable task of data analysis on such a scale. 

There are many other benefits to being able to directly use images rather than photometric features for photo-$z$ estimation, predominantly that the deep learning algorithms could extract far more information than from the magnitudes alone \citep{pasquet2019photometric, schuldt2020photometric}. Indeed previous studies such as \citet{Soo} worked hard to include morphological parameters which are implicitly contained in the images and that the deep learning algorithms could extract. Furthermore, other work has been done which found that deep learning methods have been able to produce photo-$z$ estimates which outperform the previously best performing traditional methods based on kNN or RFs \citep{d2018photometric}. However, the deep learning algorithms are often much slower and more computationally expensive to run, and there has been little investigation into whether the benefits of these methods is worthwhile in the long term. 

Here we investigate if it is worth using deep learning on images compared to traditional ML using only the magnitudes as features, applying different types of convolutional neural networks (CNNs) as well as mixed-input models which use both images and magnitudes as inputs. In Sec.~\ref{sec:Data} we describe the data collected and used to train and test the machine learning algorithms which are outlined in Sec.~\ref{sec:Methods} along with the metrics and optimisation process. We then present the results in Sec.~\ref{sec:Results} with discussions about how the mixed-input inception CNN was able to achieve such low errors and outperform all the other algorithms, before concluding in Sec.~\ref{sec:Conclusions}.

\section{Data}
\label{sec:Data}

The data used to train and test the different machine learning algorithms came entirely from the Sloan Digital Sky Survey data release 12 (SDSS DR-12) \citep{york2000sloan, alam2015eleventh}. In this work we compiled $1059678$ galaxy images with each image comprising of the five (u, g, r, i, z) wavelength bands, as well as downloading the corresponding photometric and spectroscopic data. It was a requirement to have both spectroscopy and photometry performed to be able to use the spectroscopic redshift as the true value and use the photometric magnitudes as features to compare traditional ML algorithms as well as use in the mixed-input models.

While the total number of galaxies which met the requirement of having an associated spectroscopic redshift was closer to 2 million \citep{beck2016photometric}, we decided that a training set of 1 million galaxies was sufficient. This decision was made as there wouldn't be a large difference in error performance going from 1 to 2 million galaxies. Furthermore, the scaling of the algorithms would only be visibly different with changing orders of magnitude of the training set (as we display later in figures~\ref{fig:benchmarks_training_times} - \ref{fig:benchmarks_mse_points}), and therefore simulations would have then been required to be able to include more galaxies and reach the next order of magnitude.

\begin{figure}
    \centering
    \includegraphics[scale = 0.7]{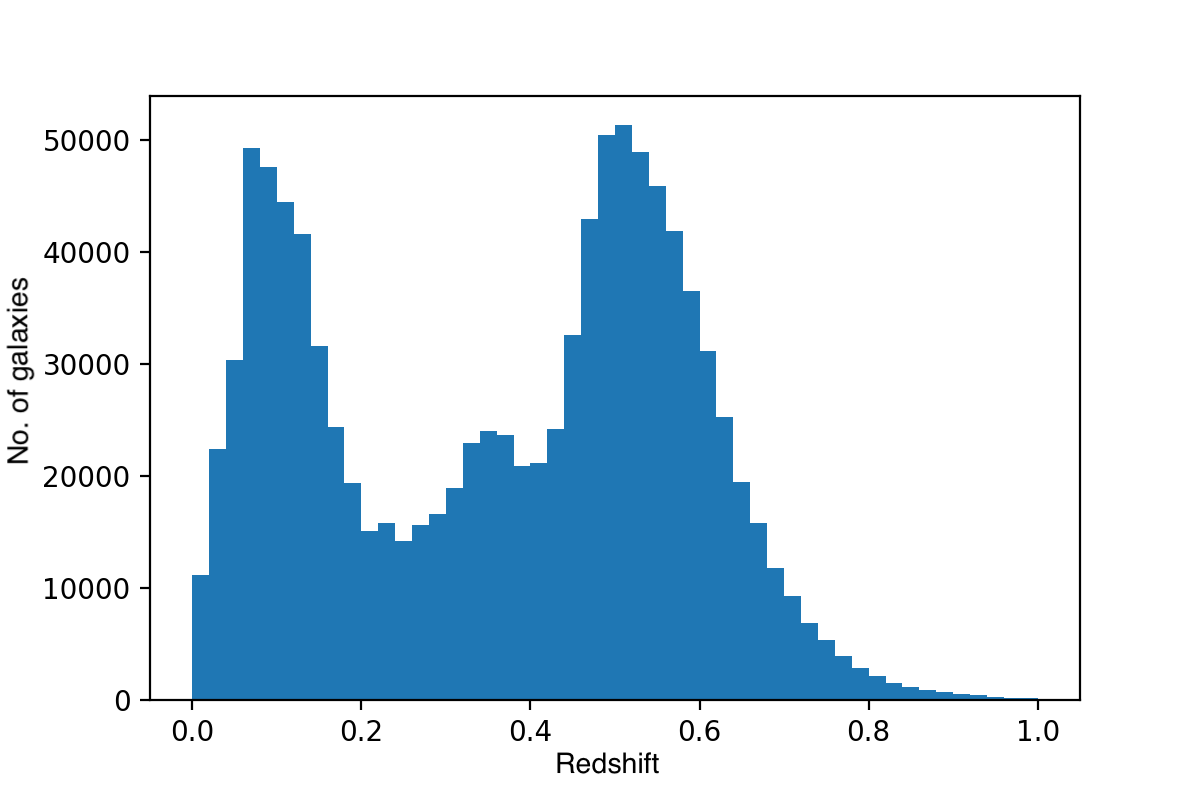}
    \caption{Plot showing the redshift distribution of the 1 million galaxies used in the study. The histogram was plotted with 50 redshift bins between $z = 0$ and $z = 1$ and displays the overall redshift distribution of galaxies in SDSS with the two peaks caused by the difference between the galaxies observed during the main galaxy survey and BOSS as described by~\citet{beck2016photometric}.}
    \label{fig:redshift_dist}
\end{figure}

The dataset used was also kept clean by requiring complete photometry where there were no missing values of any magnitudes which could have negatively impacted the redshift estimations and biased the results. Furthermore, the redshift range was kept to only include galaxies with $z < 1$, with the final distribution of galaxies used shown in figure~\ref{fig:redshift_dist}. Although this simplified the problem rather than having a larger redshift range, this distribution matched that of the overall SDSS survey as shown by \citet{beck2016photometric}, and as the data used was representative of SDSS it therefore allowed for valid estimates to be made in this range.

To generate the images used by the deep learning algorithms we first downloaded the full frames made available by SDSS which each comprised of five flexible image transport system (fits) files for the five different filters. For each galaxy within the frame we then generated $32\times32$ pixel images by centering the frame on the galaxy and cropping. We found that $32\times32$ pixel images were sufficiently large to contain the full galaxy and surrounding sky for even the closer galaxies, and allowed for much smaller file sizes when compared with $64 \times 64$ pixel images which other studies had previously tested. 

The final result was a set of five $32\times32$ pixel images for each galaxy which we saved in a numpy array \citep{harris2020array} with shape $(32\times32\times5)$ which could then be used with the deep learning python package Keras \citep{chollet2015keras}, our chosen interface for TensorFlow \citep{tensorflow2015-whitepaper}. 

\section{Methods}
\label{sec:Methods}

With the magnitude data and images downloaded and in a usable format we were then able to apply the machine learning algorithms. The algorithms tested were as follows. A convolutional neural network (CNN) and an inception module CNN which both used the image data as the input. A random forest (RF) and extremely randomised trees (ERT) that had previously been found to be the best performing traditional methods \citep{Henghes} and which only used magnitude features. And two experimental mixed-input models, which combined a CNN or inception module CNN with a multi-layer perception to use both the image data and magnitude features as inputs. Each of the algorithms is described in more detail below. 

The algorithms were trained multiple times while varying the size of the training set used (up to 1 million galaxies), and tested on a fixed test set of 59678 galaxies. By doing this we were able to benchmark the different algorithms to determine their scalability as described in Sec.~\ref{sec:benchmarking}. The optimisation process is also described in Sec.~\ref{sec:optimisation} and the metrics used to evaluate the models' performances are given in Sec.~\ref{sec:metrics}. 

\subsection{Convolutional Neural Networks}

Artificial neural networks are algorithms which aim to mimic the human brain \citep{mcculloch1943logical}. They are made up of many interconnected nodes called neurons which, similar to their biological counterparts, are used to signal and process information. This is done by taking the inputs, which can either be feature values from the data (in this case the pixel values from the images), or outputs from other neurons, and then calculating the weighted sum of of these inputs. The weights are taken from the connections between neurons and are the key property which is varied during the training process to learn the best possible network. A bias is then also added to the weighted sum which is finally passed to an activation function that produces the output.  

Neural networks are comprised of multiple layers with a minimum of an input layer, and output layer, but typically with at least one `hidden' layer of nodes in-between and in the case of deep neural networks many hidden layers are employed \citep{lecun2015deep}. The neurons of each layer are connected to some (or all in the case of fully-connected networks) of the neurons from the previous layer. Convolutional neural networks (CNNs) \citep{fukushima1982neocognitron} are a type of deep neural networks which include convolutional layers. They were also inspired by a biological counterpart in the visual cortex where neurons only respond to stimulus in a specific region called the receptive field and the receptive fields of neurons then overlap to cover the entire region \citep{hubel1968receptive}. 

\begin{figure}
    \centering
\begin{adjustwidth}{-1cm}{}
    \includegraphics[scale=0.3]{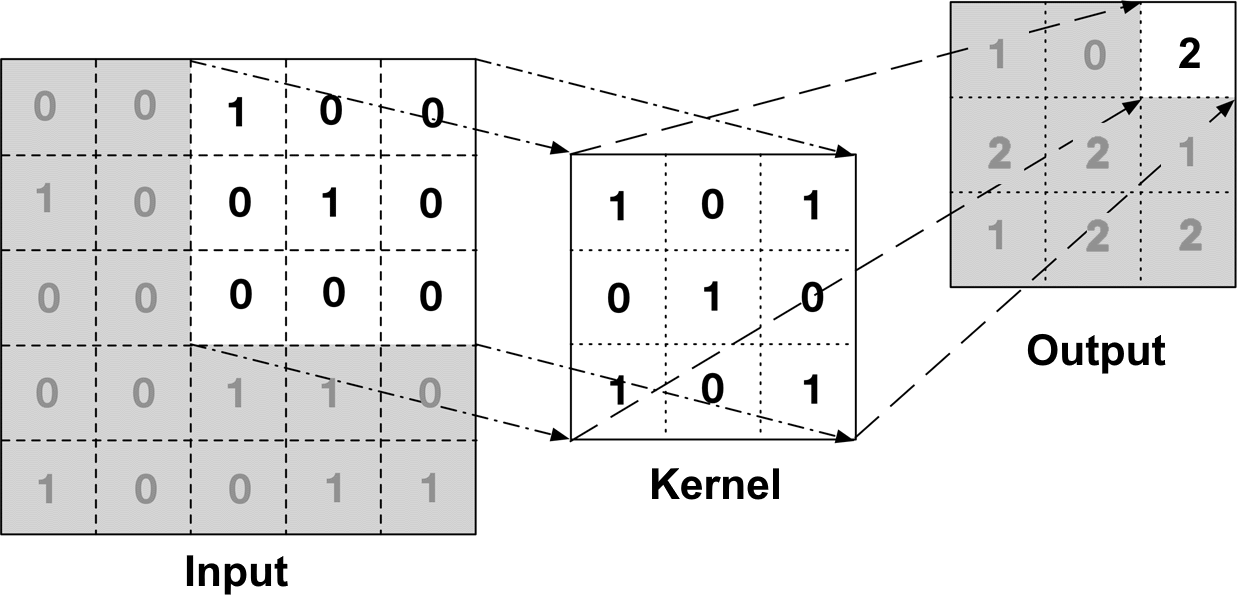}
\end{adjustwidth}
    \caption{In this figure we illustrate how a convolutional filter is applied to an input as part of a CNN. The filter (or kernel) is applied across the input, being shifted by the stride (here $=1$) each time before calculating the dot product to produce the output.  
    }
    \label{fig:conv_layer}
\end{figure}

In CNNs the convolutional layers act to convolve the input in a similar way to the visual cortex. The layer works by using filters (or kernels), which are initially random values that get updated during training to provide more relevant values. Each filter is applied across the volume of the input data as shown in figure~\ref{fig:conv_layer}, computing the dot product between the filter values and input values, and producing an activation map (or feature map). This activation map highlights regions of the input, learning features or patterns in the input images, and the maps from all filters are stacked to form the output of the convolutional layer.

As well as being able to learn features directly from the input image data, CNNs allow for much smaller overall network architectures as filters require far fewer learnable parameters compared with the equivalent fully connected network. This makes them perfectly suited to image data where large images would quickly result in exploding gradients in a traditional neural network. Furthermore, the spatial relationship between features is preserved by the convolutional layers, so for data where spatial information is important (like images) CNNs are a natural choice. 

After the convolution an activation function is used which acts to introduce nonlinearities and decides which neurons fire. The most common activation function is the rectified linear unit (ReLU) \citep{nair2010rectified}, which is defined in equation~\ref{eq:relu}. While the sigmoid function and hyperbolic tangent used to be popular choices for the activation function, ReLU sets negative values to $0$ and is much faster without decreasing accuracy. Recently an adapted parametric ReLU (PReLU) was introduced as definted in equation~\ref{eq:prelu} \citep{he2015delving}, which is an example of a `leaky' ReLU but where the coefficient of leakage, $a$, is an extra parameter that is also learned. 

\begin{equation}
\label{eq:relu}
    f(x) = \left \{ \begin{array}{rcl}0 & \mbox{for} & x < 0\\ x & \mbox{for} & x \ge 0\end{array} \right.
\end{equation}

\begin{equation}
\label{eq:prelu}
    f(x) = \left \{ \begin{array}{rcl}ax & \mbox{for} & x < 0\\ x & \mbox{for} & x \ge 0\end{array} \right.
\end{equation}

\begin{figure}
    \centering
    \includegraphics[scale=0.4]{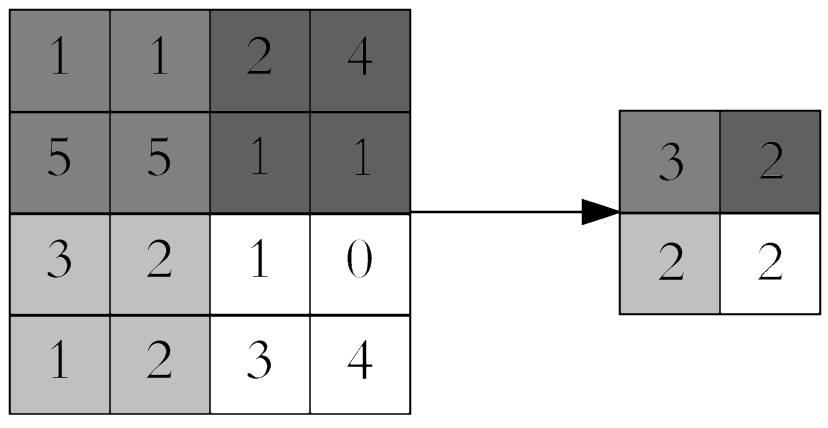}
    \caption{In this figure we show how the down-sampling is performed by a ($2\times2$) average pooling layer (with stride $= 2$), and how it acts to reduce the size of the input.}
    \label{fig:pooling}
\end{figure}

Another type of layer which is used in CNNs is the pooling layer. Pooling layers, like the one shown in figure~\ref{fig:pooling}, reduce the size of activation maps by combining the outputs within a set space (typically $2\times2$). The two types of pooling take either the maximum value (max-pooling) or average value (average-pooling) within this space and by down-sampling act to further reduce the computational demand of the CNN as well as reduce overfitting. 

The final layers which make up most CNNs are the fully connected layers. Unlike the convolutional layer the neurons of fully connected layers don't share weights and are connect to every other neuron in the previous layer. Typically these final fully connected layers behave in a similar way to a traditional neural network, taking the features extracted by the convolutional layers and further processing the information to reach the final predictions. 

The full CNN architecture used in this study is shown in figure~\ref{fig:CNN_architecture} and made use of two convolutional layers, each followed by an average pooling layer, to extract the information from the images. It then fed the flattened $4096$ long array into two dense layers, with $1024$ followed by $32$ neurons, before a final dense layer was used to give the single value output for the predicted redshift. This architecture was found to be the most effective simple CNN, and while deeper models were also tested, we found that the best performance occurred with only a few layers. 

\subsection{Inception Modules}

\begin{figure}
    \centering
    \begin{adjustwidth}{-1.2cm}{}
    \includegraphics[scale=0.28]{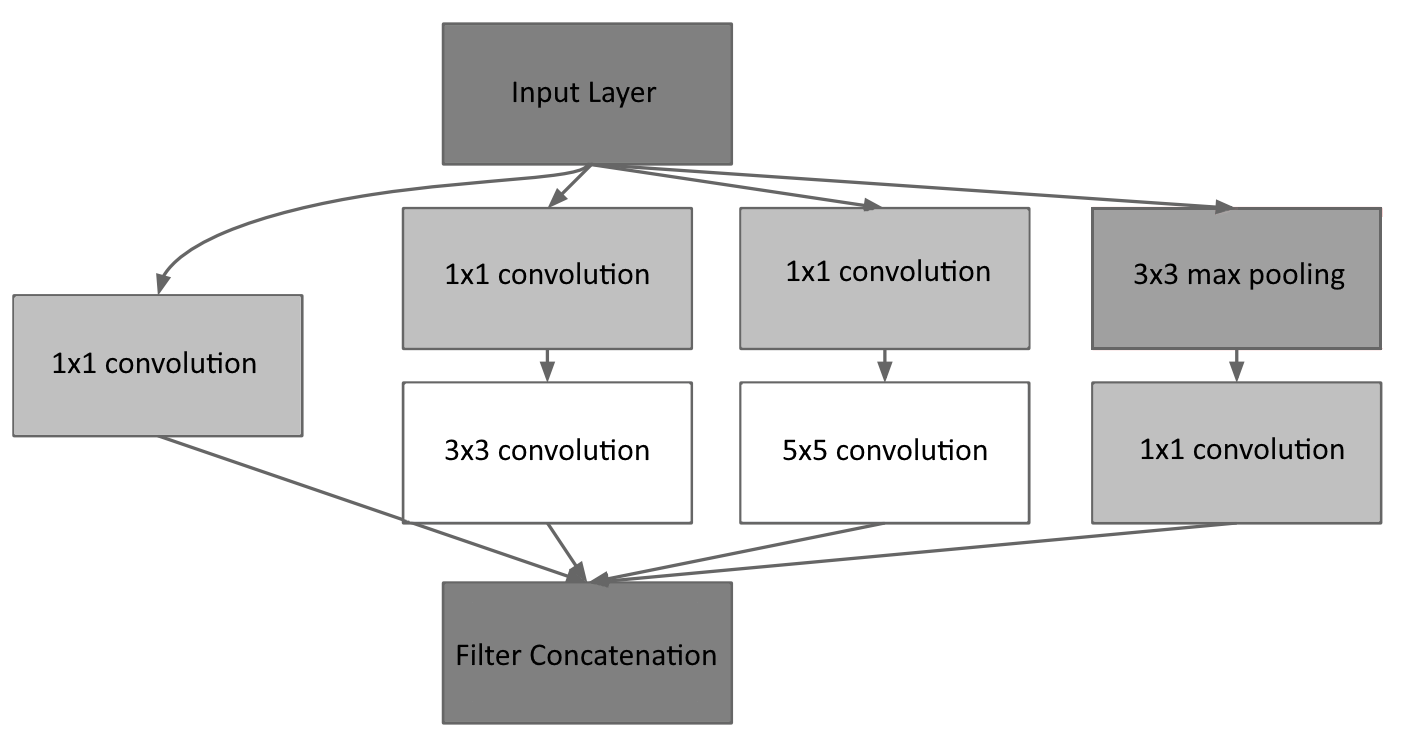}
    \end{adjustwidth}
    \caption{Figure showing a typical inception module which makes up part of the inception module CNN with each layer and kernel size labelled. The inception module is made of several convolutions, as well as a pooling layer, each applied in parallel with the ($1\times1$) convolutions acting to improve the computational efficiency. The outputs are then concatenated to produce the output of the inception module which can then be fed into the CNN similar to any other layers. }
    \label{fig:inception_module}
\end{figure}

Many variations of CNNs are possible, one example is the inception module CNN which makes use of inception modules. These are sets of convolutional layers applied in parallel rather than sequentially, as displayed in figure~\ref{fig:inception_module}. Using convolutional layers with a kernel size of $(1\times1)$ before then applying larger $(3\times3)$ or $(5\times5)$ kernels allows for more efficient computation for deeper networks \citep{inception}.

In the inception module CNN implemented here (shown in figure~\ref{fig:inception_module_cnn_architecture}), we used two different sets of inception modules. The first had the same architecture as displayed in figure~\ref{fig:inception_module}, however, after four sets of these inception modules the resulting output was too small to usefully apply a $(5\times5)$ kernel to, and instead we simply removed this layer from the inception module. Following this smaller inception module, the flattened $192$ long array was fed into two further dense layers each with $1096$ neurons before the final dense layer gave the predicted redshift. 

Unlike the simple CNNs tested, when testing different configurations of inception module CNNs we found that the deeper models performed best. As well as changing the depth of the network we also tested different inception modules, changing the size of the kernels and swapping the max pooling layer for an average pooling layer. This resulted in a somewhat different model to that used by \citet{pasquet2019photometric} while still resembling the original GoogLeNet network.

\subsection{Mixed-input Models}

It is also possible to include multiple inputs in a CNN. This works by defining separate input data and running through networks in parallel before then combining the networks by concatenating similar to how the inception modules are built. In this work we experimented with two mixed-input models which took both SDSS images with shape ($32\times32\times5$) used by the CNNs, as well as the 5 magnitudes (u, g, r, i, z) which were used as the sole features for the random forests. To handle the magnitude features a simple multi-layer perceptron (MLP) was used which included five fully connected layers each with 1024 neurons (for full network architectures used see Appendix~\ref{sec:network_architectures}). 

A perfect CNN would theoretically be able to extract all useful information from the galaxy images (including the magnitudes which are obtained from performing photometry on the same images), and hence render the mixed-input models superfluous. However, as we saw from the results, the mixed-input inception CNN was able to outperform the inception CNN it was based off where the only difference was the magnitude features also being included. This suggested that the CNN wasn't able to perfectly extract the magnitudes from the images and by explicitly providing them as additional features to be handled by a MLP we were able to boost the performance. 

In other cases of machine learning it can be ill-advised to use features which are highly correlated. This is due to the chance that the model will output results which vary drastically and wouldn't generalise to other datasets \citep{goldstein1993conditioning}. However, due to the difference in extracting information from the images through convolutional layers which result in more abstract features than the magnitudes, and the fact that the results we saw suggested the explicit inclusion of magnitude features helped rather than hurt the model performance, we concluded that this wasn't a multicollinearity issue \citep{garg2013comparison}.

There are two additional reasons for avoiding using correlated features. First, is an increase in model complexity which results in slower models than if fewer, independent features were used. However, in this case the inclusion of a MLP to handle the additional magnitude features had a minimal effect on the overall speed of the model (as can been seen in figures~\ref{fig:benchmarks_training_times}\&\ref{fig:benchmarks_inference_times_training}). The second reason is that using correlated features often makes models less interpretative. This point was also less of an issue in this case as the addition of the magnitudes as features was physically motivated and the MLP that handled these features was kept separate to the rest of the CNN. 

This allows the mixed-input models to be thought of as a combination of two separate models. This process of taking a combination of different models to give the final prediction is more widely used, and when using neural networks one has the option to implement subnetworks (as we have done here). Subnetworks, such as the CNN and MLP we used, handle the different inputs separately before their outputs are concatenated into a single feature vector which can then be used to give the final output prediction \citep{burkov2019hundred}.

\subsection{Random Forests}

Random forests (RFs) \citep{RF1} are built from many decision trees, where the predictions of all the individual trees are averaged to give the prediction of the overall forest. Each decision tree is a non-parametric algorithm which works by taking the input features and applying simple if-then-else statements which give decision boundaries to split the data into branches until eventually the data reaches a leaf node where there are no further splits. The trees are trained recursively whereby the features splits are chosen to give the most information gain (the largest difference between the two branches). 

By averaging the predictions of many individual trees, the resulting RF has a lower variance. This is also achieved by employing extra elements of randomness. First, the data used for each tree is a random subset of the overall training set \citep{Bagging}. Second, the feature splits used are taken from a random subset of the total possible feature splits rather than always taking the split which resulted in the largest information gain. These random steps result in decision trees with higher errors, but after averaging the predictions these errors cancel out and the overall RF is a much more robust model with lower variance and less overfitting. 

\subsection{Extremely Randomised Trees}

Extremely Randomised Trees (ERT) \citep{ERT} is an adaptation of the RF with an additional element of randomness. As well as using a random subset of the possible features, it also takes a random threshold for each of the features before then using the best of the these random thresholds as the decision rule. This results in a model with lower variance at the cost of a slightly greater bias, however, perhaps the biggest improvement is to the efficiency of the model. As shown by the benchmarking performed the ERT is a much faster algorithm than the RF.

For details of the RF and ERT implemented in this study we provide the hyperparameters used in appendix~\ref{sec:hyperparameters}.

\subsection{Metrics}
\label{sec:metrics}

One of the most important steps of any machine learning problem is defining sensible metrics which can be used to evaluate model performance. As we treated this as a regression problem, where we found a single value for the photometric redshift, the natural choice was to calculate the three most commonly used regression metrics: mean squared error (MSE), mean absolute error (MAE), and R squared score ($R^2$). The equations for each metric is given below for a dataset with $n$ galaxies where for the $i^{th}$ galaxy, $\hat{z}_i$ is the predicted redshift, and $z_i$ is the true spectroscopic redshift.

\begin{equation} 
\text{MSE}(z, \hat{z}) = \frac{1}{n_{\text{samples}}} \sum_{i=0}^{n_{\text{samples}}-1} (z_i - \hat{z}_i)^2
\end{equation}

\begin{equation} 
\text{MAE}(z, \hat{z}) = \frac{1}{n_{\text{samples}}} \sum_{i=0}^{n_{\text{samples}}-1} \left| z_i - \hat{z}_i \right|
\end{equation}

\begin{equation} 
R^2(z, \hat{z}) = 1 - \frac{\sum_{i=1}^{n} (z_i - \hat{z}_i)^2}{\sum_{i=1}^{n} (z_i - \bar{z})^2}
\end{equation}

As well as these standard regression metrics, there are three additional metrics which are most commonly used in photometric redshift estimation: bias - the mean separation between predition and truth, precision (also $1.48 \times$ median absolute deviation (MAD) \citep{precision}) - the expected scatter, and catastrophic outlier fraction - the fraction of predictions with an error greater than a set threshold, here set $>0.10$. The results of our tests are given in table~\ref{tab:results} with each of the metrics quoted for the various different algorithms. 

\begin{equation} 
\text{Bias} = <z_{\text{pred}} - z_{\text{spec}}>
\end{equation}

\begin{equation} 
\text{Precision} = 1.48 \times \text{median}(\frac{|z_{\text{pred}} - z_{\text{spec}}|}{1 + z_{\text{spec}}}) 
\end{equation}

\begin{equation} 
\text{Catastrophic Outlier Fraction} = \frac{N(\Delta z) > 0.10}{N_{\text{total}}}
\end{equation}

\subsection{Optimisation}
\label{sec:optimisation}

Optimisation is the process of fine tuning the hyperparameters of machine learning algorithms to give the best possible predictions. These hyperparameters are parameters that are set previous to the learning process and dictate how the algorithms create the mapping from input data to answer. For traditional methods, such as the RF and ERT, this optimisation can be done by specifying a grid of hyperparameters and iteratively testing which combination of parameters gives the best predictions. This process of testing every combination of the specified grid (called brute-force optimisation) is very slow, and instead we tested $200$ random combinations within the hyperparameter grid (called random optimisation) which gave a good estimate of the best possible parameters in much less time.

Neural networks such as the CNN, inception module CNN, and mixed-input models cannot be optimised in the same way. As well as being far more computationally expensive which would make any brute-force optimisation impossibly slow, each network has a unique architecture which changes the hyperparameters that need optimising. In this study we tested various architectures, instead of defining grids of hyperparameters to test we simply went through the hyperparameters which have the greatest impact on the models (such as the number of layers, the number of neurons in each layer, and the kernel size, stride, and padding of convolutional layers) and tested different combinations to find the best preforming models.

\subsection{Benchmarking}
\label{sec:benchmarking}

Benchmarking is the process of running a set of standardised tests to determine the relative performance of an object, in this case iteratively running the training and testing of different machine learning algorithms. Here, benchmarking was performed in a similar vein to \citet{Henghes}. We recorded the time taken throughout the machine learning process and varied the size of the training dataset to be able to compare the efficiency of the various models. By combining these measurements with the error results for the photometric redshifts we were able to better understand the performance of the different algorithms and give more discussions along with plots in section~\ref{sec:Results}.

Generally the hardware used to test each algorithm should be kept the same, however, in this case as we were comparing CNNs with traditional ML methods, the CNNs were trained using a graphics processing unit (GPU) whereas the RF and ERT were trained using the central processing unit (CPU). While this did change the nature of the comparison, it was still a valid test as both the GPU and CPU used were simply standard laptop components (an Nvidea GTX1050Ti and intel i9-8950hk). Additionally this highlights one of the key differences between deep learning methods, which are highly parallelisable, and traditional ML methods, which often aren't. Even in the case of RFs which are also parallelisable, and indeed were run over multiple CPU cores, they don't benefit to the same extent as CNNs can when run using thousands of cuda cores \citep{kirk2007nvidia}.

\begin{table*}
\centering
	\caption{Results of testing the different machine learning algorithms, where each algorithm was trained using 1000000 galaxies. The RF and ERT both used photometric data, whereas the CNN and Inception module CNN used images data, and the mixed-input CNNs used both images and photometry.}
	\label{tab:results}
	\begin{tabularx}{\textwidth}{XXXXXXX}
	\toprule
	\thead{} & \thead{Random \\ Forest \\ (RF)} & \thead{Extremely\\ Randomised \\ Trees (ERT)} &  \thead{Convolutional \\ Neural \\ Network (CNN)} & \thead{Inception \\ Module CNN \\} &  \thead{Mixed-input\\ CNN \\ } & \thead{Mixed-input\\ Inception \\ CNN} \\
	\midrule
	\addlinespace[0.2cm]
	MSE &  0.01253 & 0.01261 & 0.01009 & 0.00956 & 0.00997 & \textbf{0.00916} \\ 
	\addlinespace[0.2cm]
	\midrule
	\addlinespace[0.2cm]
	MAE &  0.05003 & 0.05067 & 0.04388 & 0.04310 & 0.04154 & \textbf{0.03966} \\
	\addlinespace[0.2cm]
	\midrule
	\addlinespace[0.2cm]
	$R^2$ & 0.76154 & 0.76002 & 0.80809 & 0.81810 & 0.81030 & \textbf{0.82567} \\
	\addlinespace[0.2cm]
	\midrule
	\addlinespace[0.2cm]
 	Bias & 0.03645 & 0.03707 & 0.03143 & 0.03090 & 0.02944 & \textbf{0.02816} \\
 	\addlinespace[0.2cm]
 	\midrule
 	\addlinespace[0.2cm]
 	Precision &  0.03076 & 0.03170 & 0.02985 & 0.02987 & 0.02764 & \textbf{0.02588} \\
 	\addlinespace[0.2cm]
 	\midrule
 	\addlinespace[0.2cm]
 	Catastrophic Outlier Fraction &  0.04722 & 0.04866 & 0.03619 & 0.03309 & \textbf{0.03048} & 0.03075\\
 	\addlinespace[0.2cm]
 	\bottomrule
	\end{tabularx}
\end{table*}

\begin{figure*}
    \centering
    \begin{adjustwidth}{-0.7cm}{}
    \includegraphics[scale = 0.57]{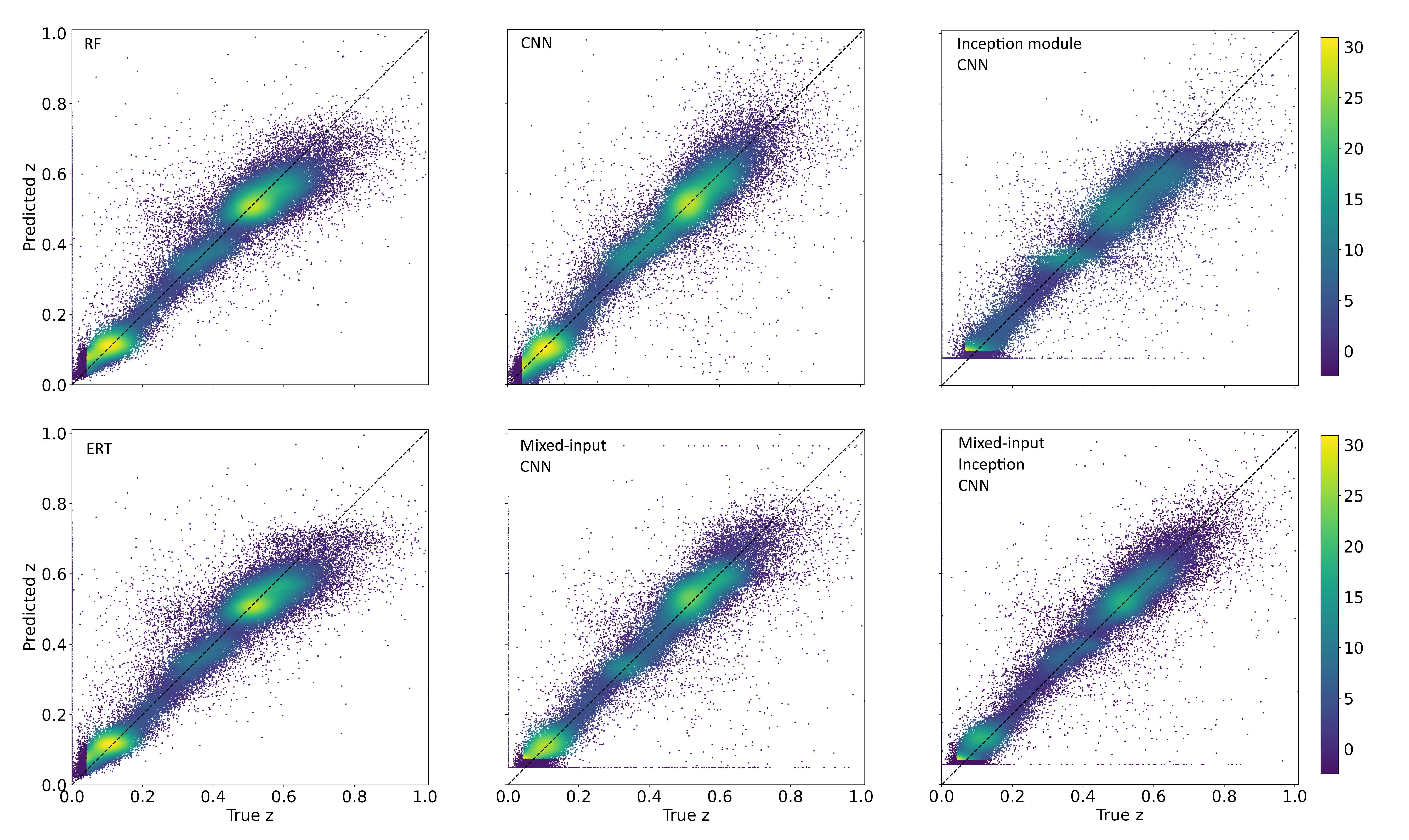}
    \end{adjustwidth}
    \caption{Density-scatter plots of the photometric redshift estimates against the true spectroscopic redshift for each of the algorithms tested.}
    \label{fig:redshifts}
\end{figure*}

\section{Results}
\label{sec:Results}

\begin{figure*}
\begin{multicols}{2}
    \begin{adjustwidth}{-0.7cm}{}
    \includegraphics[scale=0.34]{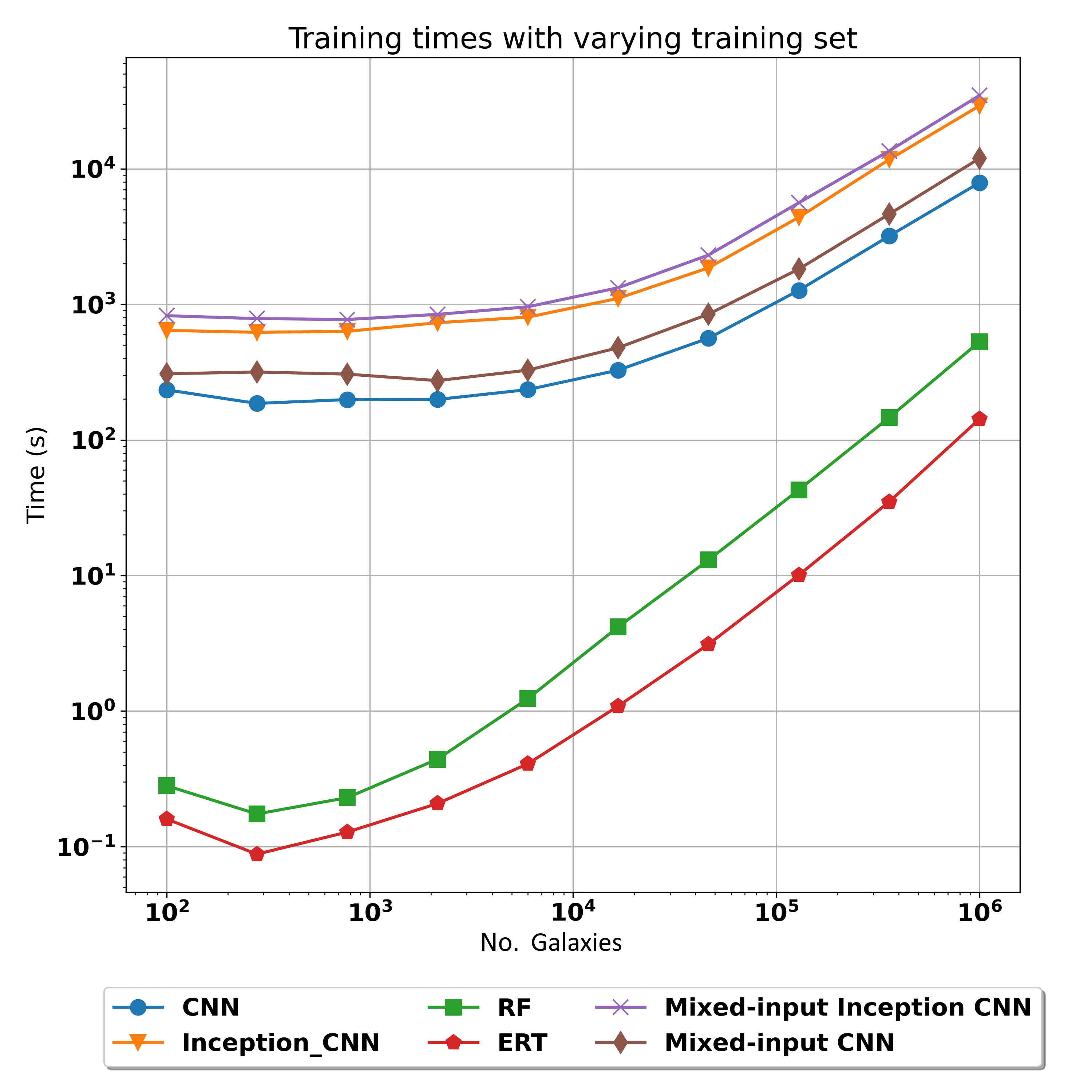}\par 
    \end{adjustwidth}
        \caption{Plot showing how the training time changes with the number of galaxies used in the training set to display how each algorithm scaled.}
        \label{fig:benchmarks_training_times}
    \includegraphics[scale=0.34]{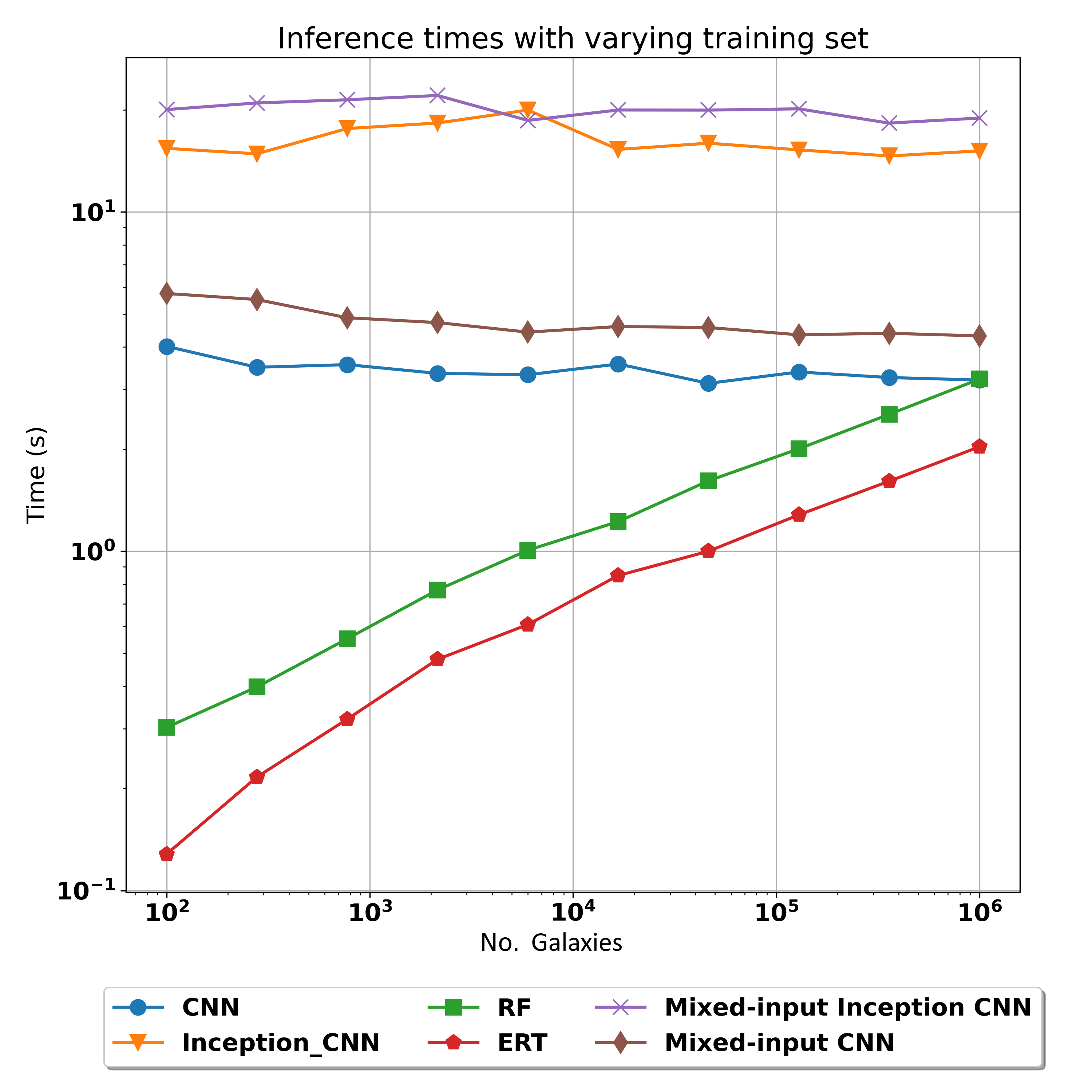}\par 
        \caption{Plot showing how the inference time changes with the number of galaxies used in the training set to display how each algorithm scaled.}
        \label{fig:benchmarks_inference_times_training}
    \end{multicols}
\begin{multicols}{2}
    \begin{adjustwidth}{-0.7cm}{}
    \includegraphics[scale=0.34]{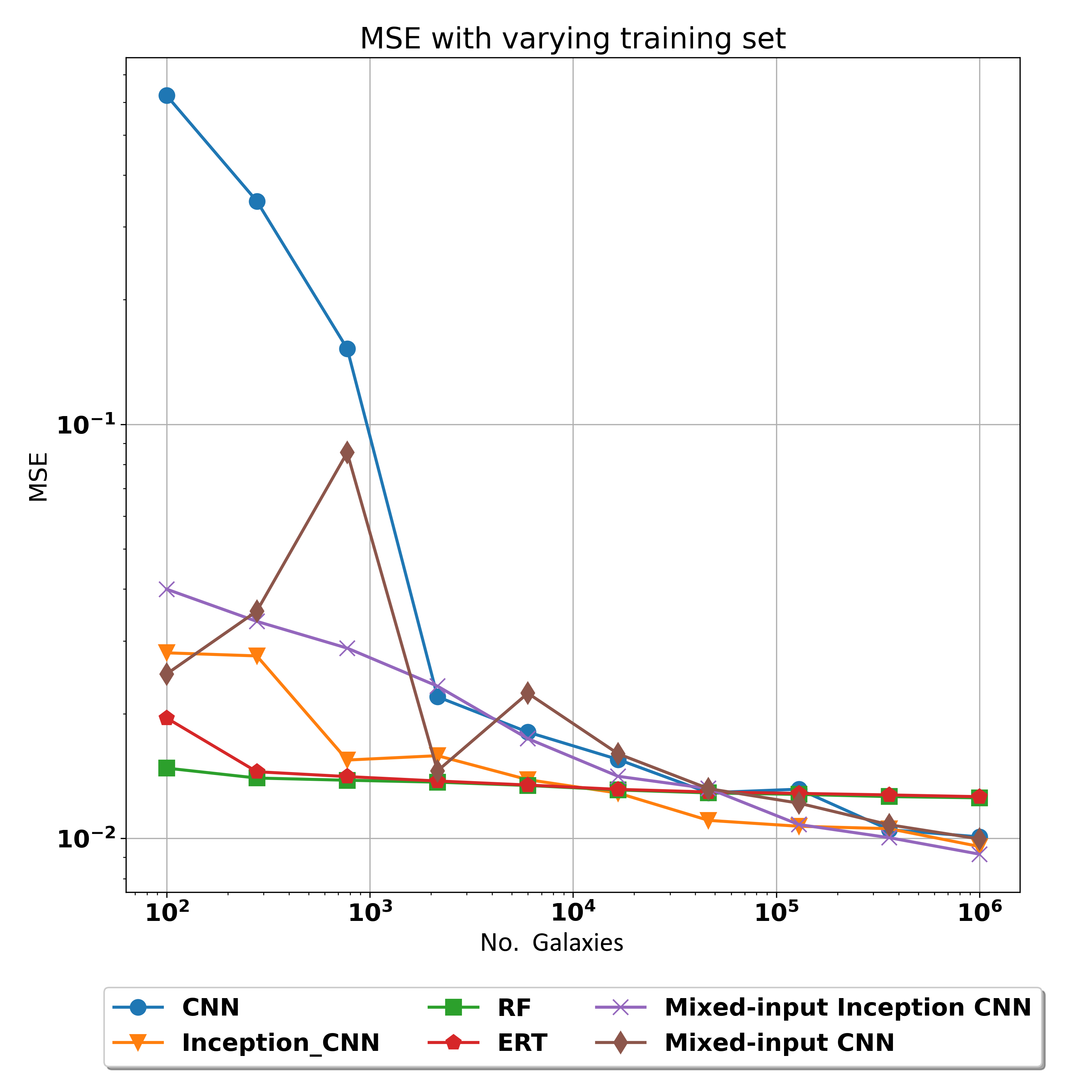}\par
    \end{adjustwidth}
        \caption{Graphs of the Mean Squared Error (MSE) plotted against the number of galaxies in the training set to show how each algorithm's performance scaled.}
        \label{fig:benchmarks_mse_points}
    \includegraphics[scale=0.34]{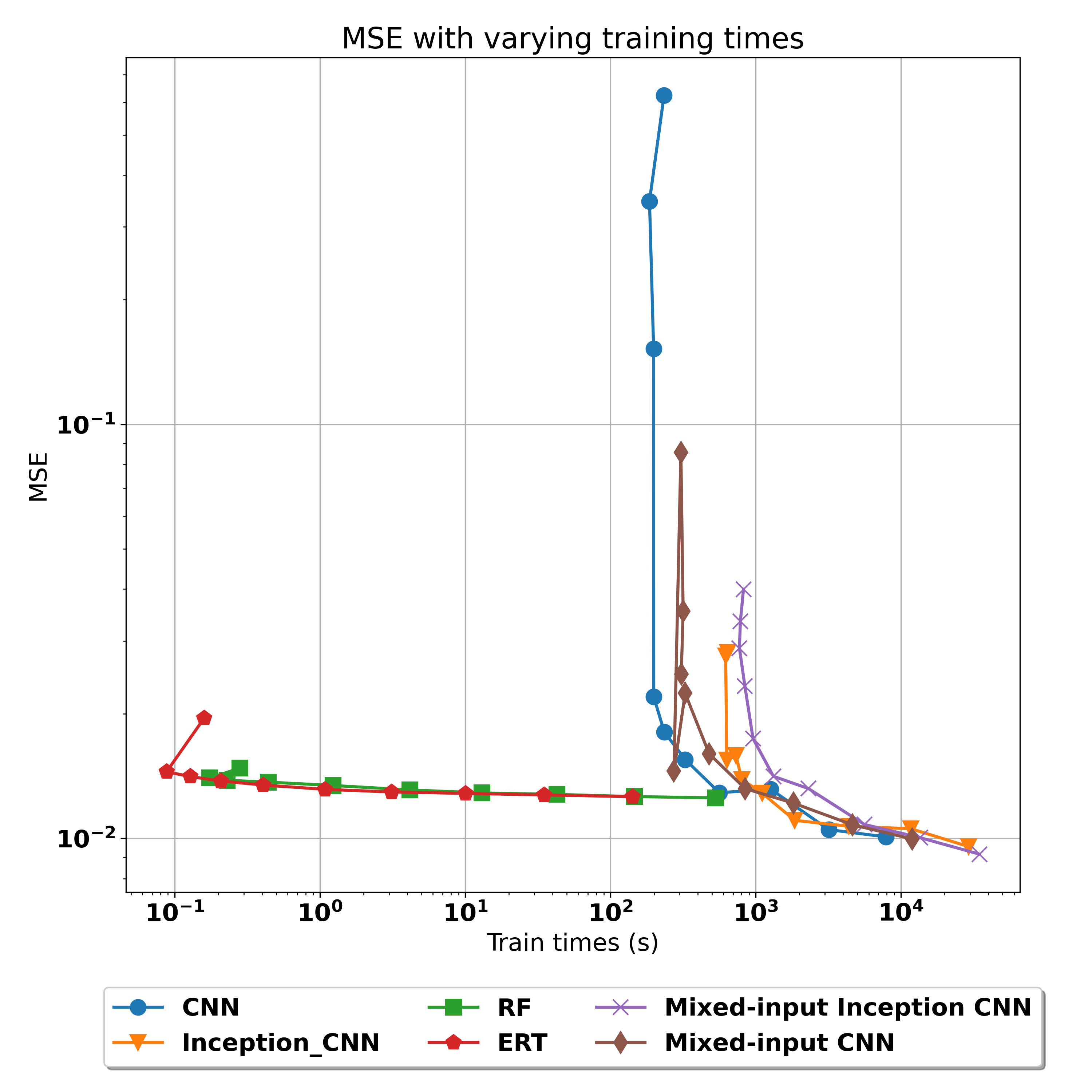}\par
        \caption{Graphs of the Mean Squared Error (MSE) plotted against the time taken to train the algorithms. }
        \label{fig:benchmarks_mse_time}
\end{multicols}
\end{figure*}

The results of our investigation are presented in multiple ways. Table~\ref{tab:results} details the error metrics achieved by each of the machine learning algorithms when using the full 1 million galaxies in the training set. We plotted density-scatter graphs of the predicted photo-$z$ estimates against the true spectroscopic redshift in figure~\ref{fig:redshifts}, as well as plotting the results of the benchmarks in figures~\ref{fig:benchmarks_training_times}-\ref{fig:benchmarks_mse_time} to show how the different algorithms scaled. Finally, the results of running the same algorithms on a smaller redshift range of $z < 0.3$ are discussed in section~\ref{sec:small_z}.

From these results we saw that the mixed-input inception module CNN was the best performing algorithm in terms of errors with a mean squared error $(MSE) = 0.009$. It also had the best performance in every other metric other than catastropic outlier fraction, where the mixed-input (standard) CNN had a slightly better fraction with both having just over $3\%$ outliers. 

While this was impressive performance it did come at a cost of being the slowest algorithm tested, which made sense being the most complex model. As seen from figures~\ref{fig:benchmarks_training_times} \& \ref{fig:benchmarks_inference_times_training} the two inception module based CNNs were the slowest algorithms, with the mixed-input model being only slightly slower in both training and inference than the image only model. Similarly the mixed-input CNN was only slightly slower compared with the CNN, showing that the addition of the magnitude features to the image based CNNs means only a small increase in computational requirements. 

We also saw that the traditional machine learning methods, RF and ERT, were significantly faster but also worse in terms of their error performance. While these methods could still be very useful in the absence of image data, the improvements seen by making use of CNNs make them an exciting option. Furthermore, by directly using the image data rather than the magnitude features, one could offset the increased time required to train the algorithms with the time saved due to not needing to previously extract features. 

What's more, the RF and ERT also showed the worst scaling of all the algorithms, slowing down at a faster rate as the number of galaxies included in the training set was increased. Past 1 million galaxies in the training set the RF was already slower in inference than the CNN, and with large enough datasets it's possible that they could become almost as slow during training. If for the largest datasets CNNs become faster than traditional methods then their main setback of being slower and more computationally expensive would no longer be of concern. 

The performance of the RF and ERT highlighted the improvements possible when including image data rather than using magnitudes alone, with a reduction in errors of around $25\%$. The experimental mixed-input models also showed good potential to further improve performance, however, as the inception module CNN performed better than the mixed-input CNN, it was clear that the CNN network architecture had a greater impact than the addition of the magnitudes as extra features. The improvement from inception module CNN to the mixed-input inception CNN was also much less than the improvement from RF or ERT to the CNNs (the improvement from including images), with a further error reduction at just over $4\%$.

\begin{table*}
\centering
	\caption{Results of testing the different machine learning algorithms for the redshift range $z < 0.3$. Each algorithm was retrained using 400000 galaxies with the RF and ERT both using photometric data, whereas the CNN and Inception module CNN used images, and the mixed-input CNNs used both images and photometry.}
	\label{tab:results_small_z}
	\begin{tabularx}{\textwidth}{XXXXXXX}
	\toprule
	\thead{} & \thead{Random \\ Forest \\ (RF)} & \thead{Extremely\\ Randomised \\ Trees (ERT)} &  \thead{Convolutional \\ Neural \\ Network (CNN)} & \thead{Inception \\ Module CNN \\} &  \thead{Mixed-input\\ CNN \\ } & \thead{Mixed-input\\ Inception \\ CNN} \\
	\midrule
	\addlinespace[0.2cm]
	MSE &  0.00140 & 0.00162 & 0.00072 & 0.00070 & 0.00169 & \textbf{0.00069} \\ 
	\addlinespace[0.2cm]
	\midrule
	\addlinespace[0.2cm]
	MAE &  0.02472 & 0.02855 & 0.01851 & 0.01773 & 0.02785 & \textbf{0.01693} \\
	\addlinespace[0.2cm]
	\midrule
	\addlinespace[0.2cm]
	$R^2$ & 0.76004 & 0.72196 & 0.87732 & 0.88074 & 0.71023 & \textbf{0.88230} \\
	\addlinespace[0.2cm]
	\midrule
	\addlinespace[0.2cm]
 	Bias & 0.02197 & 0.02531 & 0.01650 & 0.01573 & 0.02442 & \textbf{0.01506} \\
 	\addlinespace[0.2cm]
 	\midrule
 	\addlinespace[0.2cm]
 	Precision &  0.02218 & 0.02803 & 0.01875 & 0.01691 & 0.02631 & \textbf{0.01543} \\
 	\addlinespace[0.2cm]
 	\midrule
 	\addlinespace[0.2cm]
 	Catastrophic Outlier Fraction &  0.02245 & 0.02393 & \textbf{0.00557} & 0.00659 & 0.02338 & 0.00816 \\
 	\addlinespace[0.2cm]
 	\bottomrule
	\end{tabularx}
\end{table*}

\begin{figure*}
    \begin{adjustwidth}{-0.7cm}{}
    \centering
    \includegraphics[scale = 0.57]{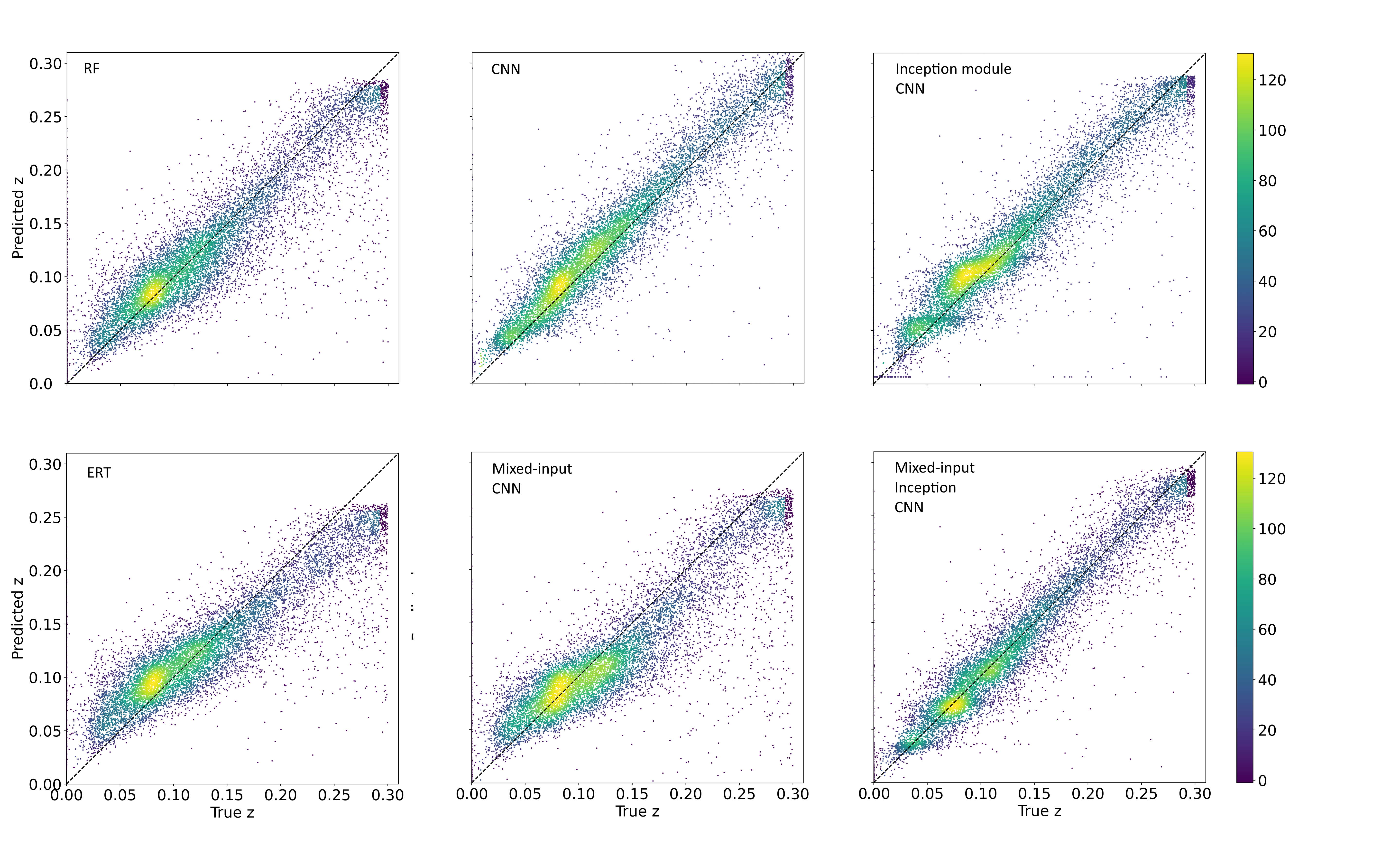}
    \end{adjustwidth}
    \caption{Graphs of photometric redshift estimates against the true spectroscopic redshift for machine learning algorithms retrained on galaxies within the range $z < 0.3$.}
    \label{fig:redshift_smaller_z}
\end{figure*}

\subsection{Lower redshift range}
\label{sec:small_z}

Although the algorithms performed well over the entire dataset we wanted to also test the performance for a smaller region to be able to more directly compare with other studies (such as \citet{pasquet2019photometric}) and see how much better the performance could be when the problem of estimating redshifts was made easier by only considering the range $z < 0.3$. 

The exact same process was carried out using the same six algorithms algorithms and the results from the retrained machine learning algorithms are given in table~\ref{tab:results_small_z}. We also plotted the redshift estimations against the true spectroscopic redshift in figure~\ref{fig:redshift_smaller_z}. 

From these we saw that in general the algorithms performed much better, reaffirming the fact that photometric redshift estimation becomes an easier problem over a shorter range. The mixed-input inception CNN continued to be the best performing algorithm with a $MSE = 0.0007$, however, there was far less separating the CNN and inception module CNN in the smaller redshift range. In fact, the CNN performed better than every other algorithm when it came to the catastrophic outlier fraction with only $0.56\%$ outliers and one of the best constrained scatter plots. 

The only algorithm which didn't show the same improvements was the mixed-input (standard) CNN, which, when it came to the smaller redshift range, performed more similarly to the RF and ERT. As we saw the mixed-input inception model perform very well it seemed like there was still potential to make use of both images and photometry, however, by not optimising the algorithms for the new dataset and using the exact same network that was used for the entire redshift range, the mixed-input CNN clearly didn't transfer as well to the new data.  

There was also a greater disparity between the other CNN based methods and traditional methods for this smaller redshift range. While there was around a $30\%$ improvement going from the RF and ERT to CNNs over the entire dataset, this increased to $50\%$ for the smaller redshift range. The image based CNNs were therefore able to provide even more advantage in this range, suggesting that the additional information extracted from the images is even more beneficial in the smaller range, possible due to the fact that the galaxies would generally occupy a larger region of the image.

However, this boost in performance for the redshift range of $z < 0.3$ also highlighted a key failing of the algorithms, in that an ideal model would generalise well enough to perform just as well across the entire redshift range. This might not be realistic as by removing a large section of the data there was far less chance of having catastrophic outliers, but it did show the benefit of splitting the data into smaller sections, and it would be beneficial to have multiple algorithms for different redshift ranges.

\section{Conclusions}
\label{sec:Conclusions}

Processing accurate photometric redshift estimations will remain a vital task of cosmological analyses. Future surveys such as LSST aim to observe more galaxies than ever before, and it is of upmost importance that the methods developed and implemented are both effective and efficient. 

Here we have shown how image-based CNN methods compare to traditional tree-based methods which make use of magnitude features from photometry. We found that the additional information the CNNs were able to extract directly from the images of galaxies allowed for a significant reduction in errors. However, as the CNNs were more complex than the RF and ERT algorithms, they were also much slower to run and required far more computational resources. 

Our results showed that the experimental mixed-input models in particular had great potential for photo-$z$ estimation. Using 1 million images of galaxies to train the algorithm the mixed-input inception CNN was able to achieve a $MSE = 0.009$. Furthermore, when the problem was simplified to only include galaxies in the range $z < 0.3$, the model achieved an even more impressive $MSE = 0.0007$, outperforming the traditional RF by $ >50\%$.

Further work would include using even more data with tens or hundreds of millions of galaxies and images (which would require the use of large scale simulations and more powerful computer architectures). The use of more powerful CPUs and GPUs in high perfomance computing systems could allow for better practices in benchmarking and set a standard system. Additionally, by stretching the amount of data further we could then determine with certainty at what point the CNNs would become faster than the RF and ERT, as well as discover whether increasing the amount of data used in the training set would eventually have no effect on model performance.

\section*{Acknowledgements}

B.H. was supported by the STFC UCL Centre for Doctoral Training in Data Intensive Science (grant No. ST/P006736/1). 
O.L. thanks All Souls College, Oxford for a Visiting Fellowship.
Authors also acknowledge the support from following grants:  O.L.'s European Research Council Advanced Grant (TESTDE FP7/291329), STFC Consolidated Grants (ST/M001334/1 and ST/R000476/1),
J.T.'s UKRI Strategic Priorities Fund (EP/T001569/1), particularly the AI for Science theme in that grant and the Alan Turing Institute, Benchmarking for AI for Science at Exascale (BASE), EPSRC ExCALIBUR Phase I Grant (EP/V001310/1).

Funding for SDSS-III has been provided by the Alfred P. Sloan Foundation, the Participating Institutions, the National Science Foundation, and the U.S. Department of Energy Office of Science. The SDSS-III web site is http://www.sdss3.org/.

SDSS-III is managed by the Astrophysical Research Consortium for the Participating Institutions of the SDSS-III Collaboration including the University of Arizona, the Brazilian Participation Group, Brookhaven National Laboratory, Carnegie Mellon University, University of Florida, the French Participation Group, the German Participation Group, Harvard University, the Instituto de Astrofisica de Canarias, the Michigan State/Notre Dame/JINA Participation Group, Johns Hopkins University, Lawrence Berkeley National Laboratory, Max Planck Institute for Astrophysics, Max Planck Institute for Extraterrestrial Physics, New Mexico State University, New York University, Ohio State University, Pennsylvania State University, University of Portsmouth, Princeton University, the Spanish Participation Group, University of Tokyo, University of Utah, Vanderbilt University, University of Virginia, University of Washington, and Yale University.
\section*{Data Availability}

The data used in this paper came entirely from the Sloan Digital Sky 
Survey data release 12 (SDSS-DR12), and is openly available from: \url{https://www.sdss.org/dr12/}.


\bibliographystyle{mnras}
\bibliography{bib} 



\appendix

\section{Network Architectures}
\label{sec:network_architectures}
In the following pages we present plots of the full network architectures used in the four CNN-based methods. 

\begin{figure*}
\begin{multicols}{2}
    \begin{adjustwidth}{-0.0cm}{}
    \includegraphics[scale=0.12]{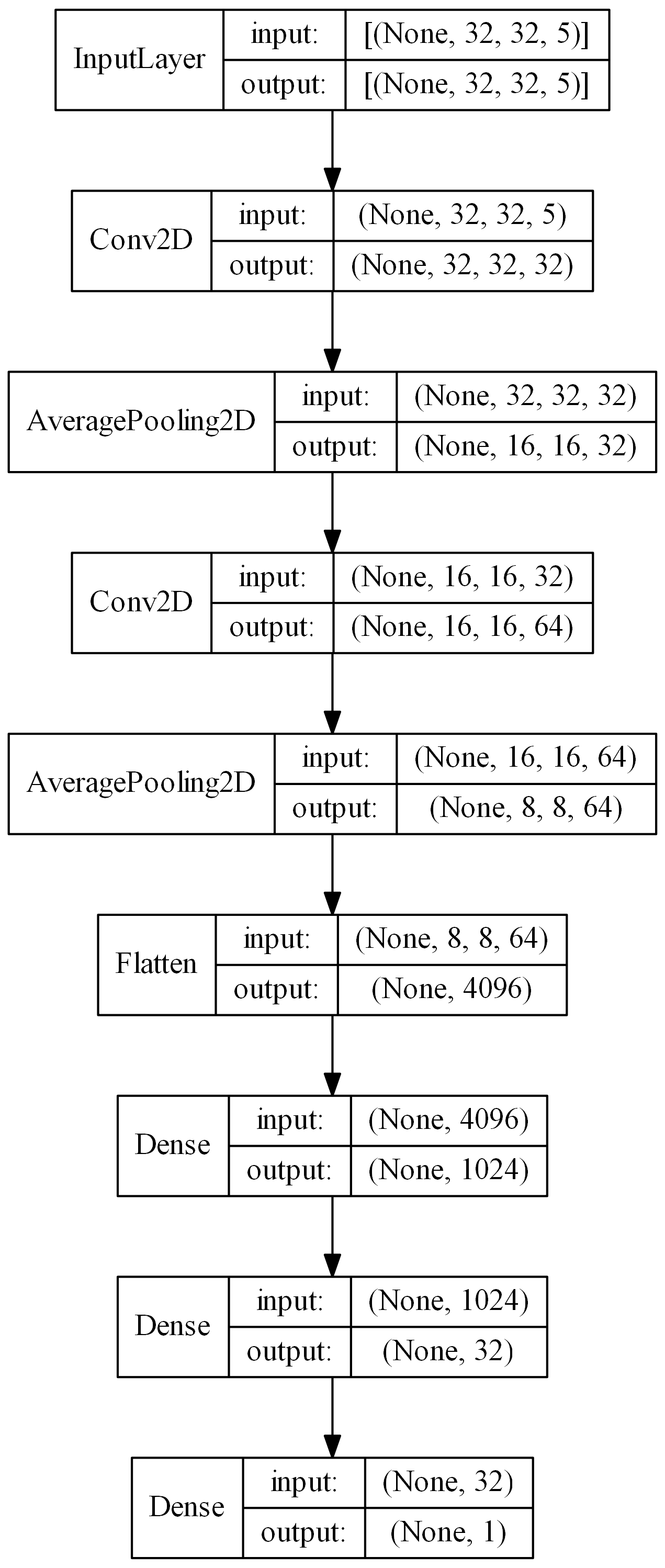}
    \end{adjustwidth}
        \caption{Network architecture of the base CNN tested. The CNN was constructed with two convolutional layers, each followed by an average pooling layer to reduce the dimensionality, before the feature map was flattened to give a 1D feature vector. This could then be handed to the two dense layers (which are the fully connected layers) that process the features before the final, single neuron layer is used to give the value of the predicted redshift.}
        \label{fig:CNN_architecture}
        
    \includegraphics[scale=0.12]{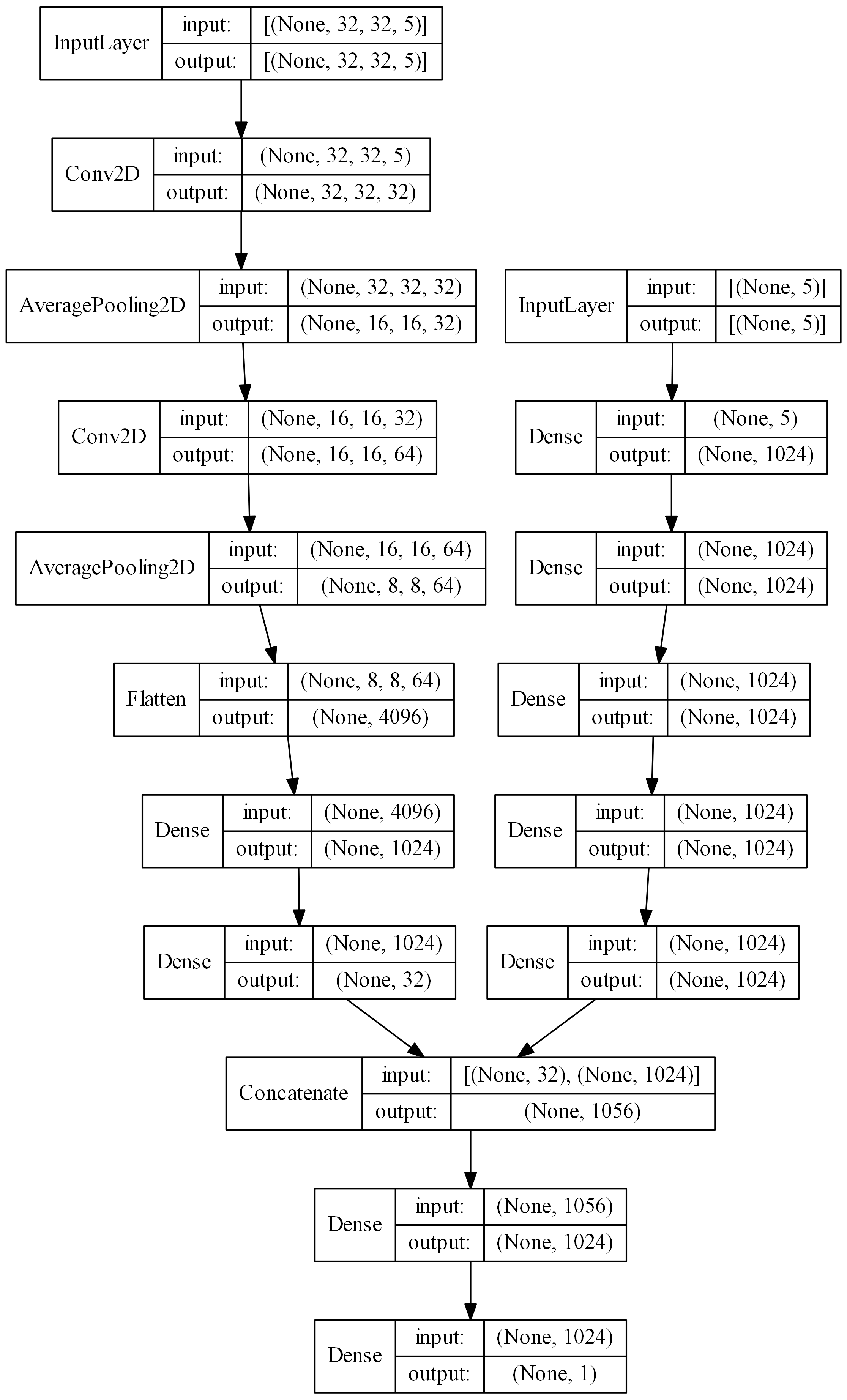}
        \caption{Network architecture of the mixed-input CNN. This model used the same CNN as \ref{fig:CNN_architecture} to handle the images, and added a MLP with 5 fully connected layers each of 1024 neurons to handle the magnitude data. The outputs of both were then concatenated before being handed to a fully connected layer and finally the single neuron layer which gave the value of the predicted redshift.}
        \label{fig:mixed_input_CNN_architecture}
    \end{multicols}
\end{figure*}

\begin{figure*}
    \centering
    \includegraphics[scale=0.070]{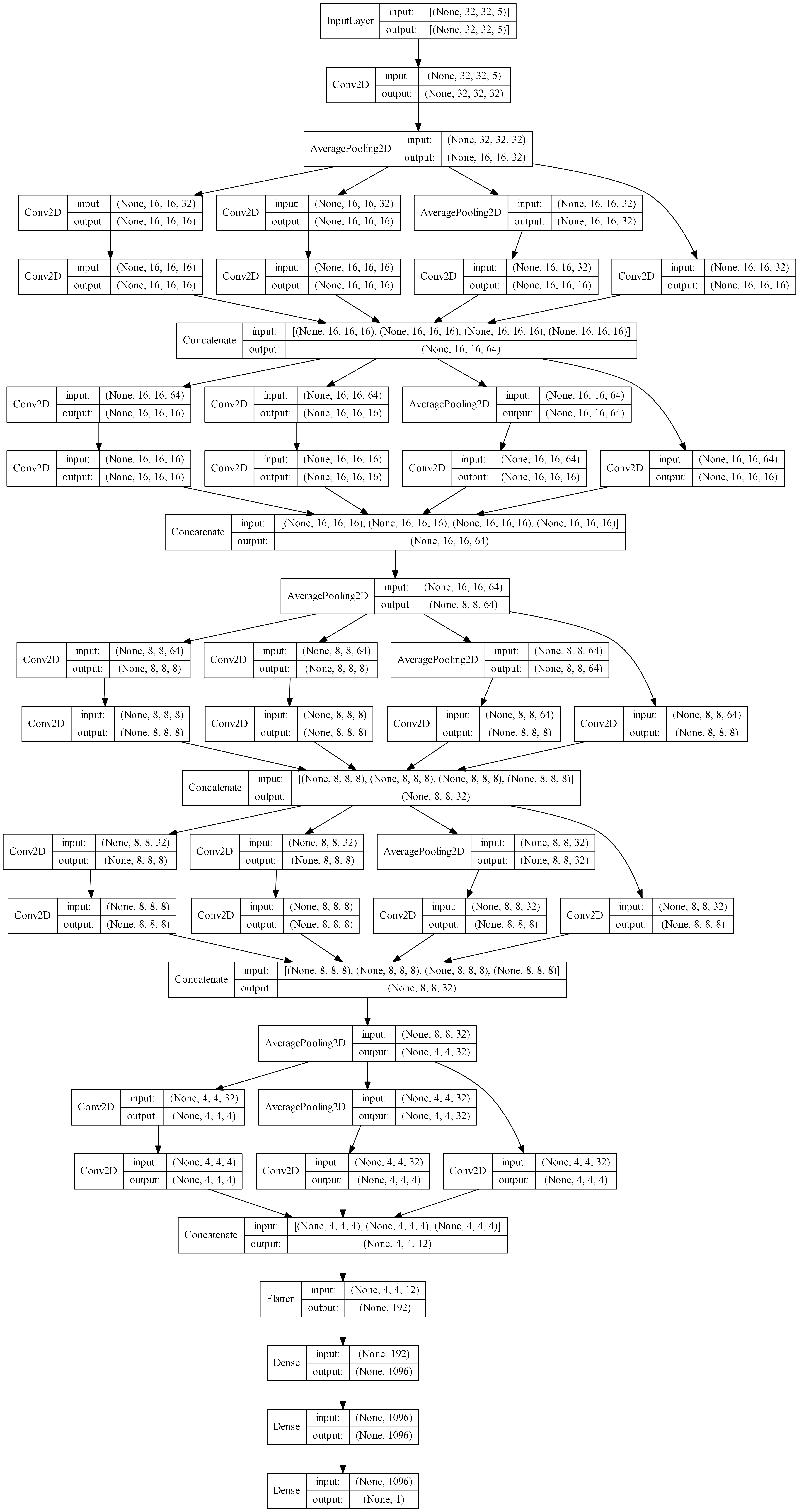}
    \caption{Network architecture of the inception module CNN. This model used a single convolutional layer and average pooling layer before applying 5 inception modules, where the fifth inception module was a modified version to be smaller and not include a $(5\times5)$ kernel. Following the inception modules the output was flattened to give the feature vector which was processed by two fully connected layers with 1096 neurons, and finally the single neuron layer to give the predicted redshift. }
    \label{fig:inception_module_cnn_architecture}
\end{figure*}

\begin{figure*}
    \centering
    \includegraphics[scale=0.070]{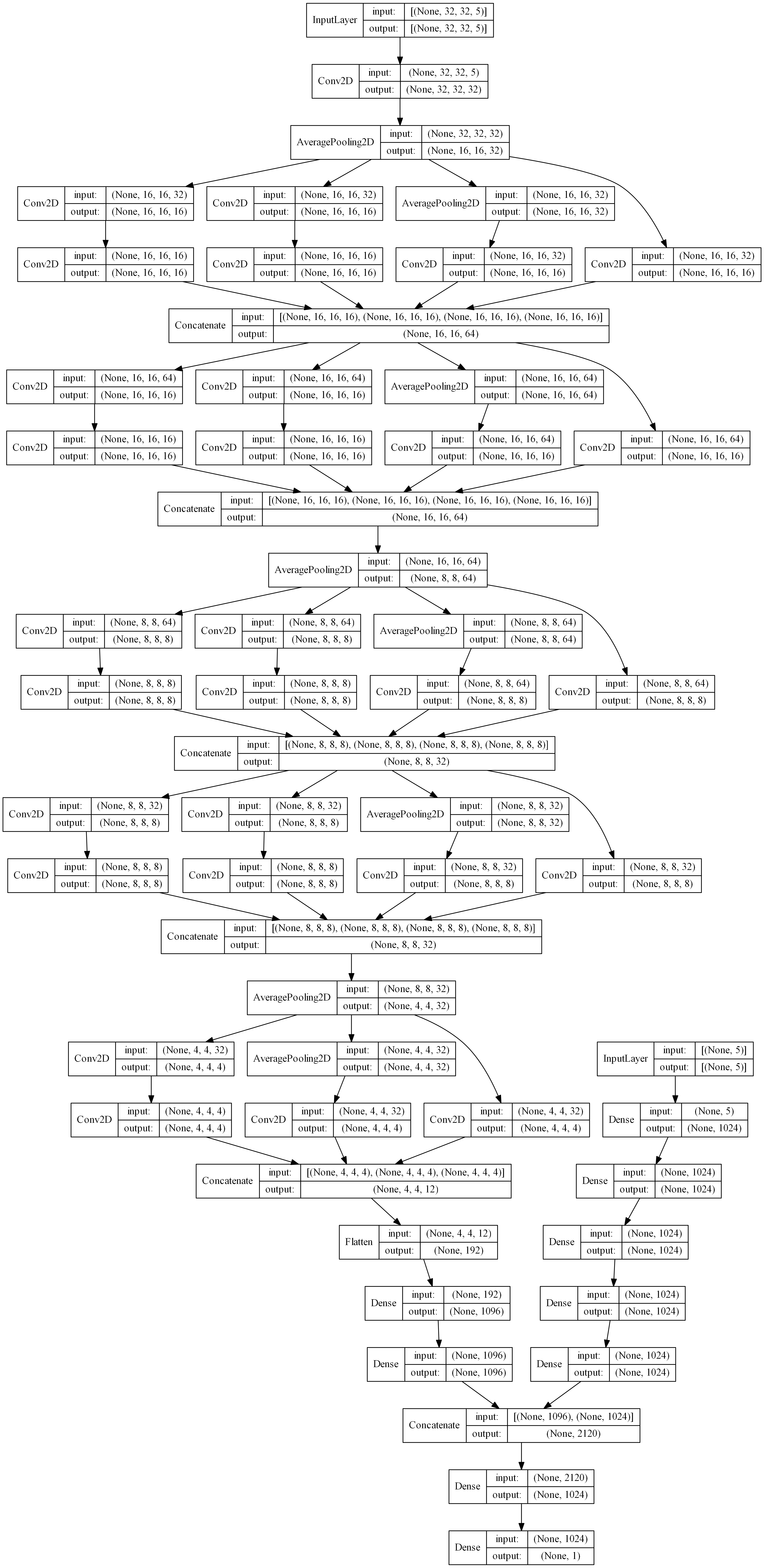}
        \caption{Network architecture of the mixed-input inception CNN. This model used the same inception module CNN as \ref{fig:inception_module_cnn_architecture} to handle the images, and added the same 5 layer MLP which was used in \ref{fig:mixed_input_CNN_architecture} to handle the magnitude features. The outputs of both were concatenated and handed to a single fully connected layer before the final single neuron layer gave the predicted redshift.}
    \label{fig:mixed_input_inception_module_cnn_architecture}
\end{figure*}

\section{Hyperparameters}
\label{sec:hyperparameters}
Here we present a table detailing the hyperpameters used by the RF and ERT methods. 

\begin{table}
\centering
\caption{Grids of hyperparameters that were used in the RF and ERT, selected by the random optimisation.}
	\label{tab:hyperparameters}
	\begin{tabular}{ccc}
		\hline
		 Classifier & Hyperparameter & Selected Value \\
        \hline
		RF & ``no. estimators" & \textbf{200}\\
			& ``max. features" & \textbf{2} \\
            & ``min. samples leaf" & \textbf{7}\\
            & ``min. samples split" & \textbf{3} \\
		    & ``min weight fraction leaf" & \textbf{0} \\
		    & ``criterion" & \textbf{mse}\\
        \hline
		ERT & ``no. estimators" & \textbf{147}\\
			& ``max. features" & \textbf{4} \\
            & ``min. samples leaf" & \textbf{3}\\
            & ``min. samples split" & \textbf{87} \\
		    & ``min weight fraction leaf" & \textbf{0} \\
		    & ``criterion" & \textbf{mse}\\
        \hline
	\end{tabular}
\end{table}


\bsp	
\label{lastpage}
\end{document}